\documentclass[preprint,prb]{revtex4}

\usepackage{graphicx}
\usepackage{dcolumn}
\usepackage{float}
\usepackage{amsmath}

\newcommand{\comment}[1]{}

\begin{document}

\title{\boldmath Doping-dependent superconducting physical quantities of K-doped BaFe$_2$As$_2$ obtained through infrared spectroscopy \unboldmath}

\author{Seokbae Lee$^1$} \author{Yu-Seong Seo$^1$} \author{Seulki Roh$^1$} \author{Dongjoon Song$^{2,3}$} \author{Hiroshi Eisaki$^2$} \author{Jungseek Hwang$^{1}$} \email{jungseek@skku.edu}

\affiliation{$^1$Department of Physics, Sungkyunkwan University, Suwon, Gyeonggi-do 16419, Republic of Korea\\ $^2$National Institute of Advanced Industrial Science and Technology, Tsukuba 305-8568, Japan\\ $^3$Department of Physics and Astronomy, Seoul National University, Seoul 08826, Republic of Korea}

\date{\today}

\begin{abstract}

We investigated four single crystals of K-doped BaFe$_2$As$_2$ (Ba-122), Ba$_{1-x}$K$_x$Fe$_2$As$_2$ with $x =$ 0.29, 0.36, 0.40, and 0.51, using infrared spectroscopy. We explored a wide variety of doping levels, from under- to overdoped. We obtained the superfluid plasma frequencies ($\Omega_{\mathrm{sp}}$) and corresponding London penetration depths ($\lambda_{\mathrm{L}}$) from the measured optical conductivity spectra. We also extracted the electron-boson spectral density (EBSD) functions using a two-parallel charge transport channel approach in the superconducting (SC) state. From the extracted EBSD functions, the maximum SC transition temperatures ($T_c^{\mathrm{Max}}$) were determined using a generalized McMillan formula and the SC coherence lengths ($\xi_{\mathrm{SC}}$) were calculated using the timescales encoded in the EBSD functions and reported Fermi velocities. We identified some similarities and differences in the doping-dependent SC quantities between the K-doped Ba-122 and the hole-doped cuprates. We expect that the various SC quantities obtained across the wide doping range will provide helpful information for establishing the microscopic pairing mechanism in Fe-pnictide superconductors. \\ \\

\noindent *Correspondence to [email: jungseek@skku.edu].

\end{abstract}

\maketitle

\section{Introduction}

Since the discovery of novel high-temperature superconductors, Fe-pnictides \cite{kamihara:2006,kamihara:2008}, intensive investigations have been performed to reveal the microscopic pairing mechanism for superconductivity\cite{ding:2008,wu:2010,mazin:2010a,basov:2011,fernandes:2022}. However, till date, the microscopic superconducting (SC) mechanism has not been elucidated. The phase diagram of Fe-pnictides is quite similar to that of copper oxide superconductors (or cuprates) \cite{mazin:2010a,basov:2011}, even though the underlying electronic ground states are different \cite{schafgans:2012,wang:2012}. Therefore, it has been suggested that these two types of high-temperature SC material systems (Fe-pnictides and cuprates) may share a microscopic pairing mechanism. These two material systems are known as correlated electron systems \cite{kotlier:2006,qazilbash:2009,weber:2010}. Remarkably, the origin of the correlations in these two material systems may not be the same; the parent compounds of cuprates are known as the Mott insulators, whereas those of Fe-pnictides are known as the Hund's metals \cite{schafgans:2012,wang:2012c}.

The electron-boson spectral density (EBSD) functions are known to contain information on the correlations between itinerant electrons through exchange of force-mediating bosons. The EBSD functions of the cuprates have been extracted from the measured spectra through various experimental techniques \cite{carbotte:2011}. Optical spectroscopy has significantly contributed to the investigation of the EBSD functions of cuprates \cite{puchkov:1996,carbotte:1999,hwang:2004,dordevic:2005,hwang:2006,hwang:2007,heumen:2009,hwang:2018}
using well-established analytical methods \cite{allen:1971,shulga:1991,sharapov:2005,dordevic:2005,schachinger:2006}. The Fe-pnictides have multiple bands at the Fermi level \cite{ding:2008,subedi:2008}, whereas the cuprates have a single band at the Fermi level \cite{dessau:1993}. Because of these different properties, in principle, the well-established method used to analyze the optical spectra of cuprates cannot be directly applied to multiband Fe-pnictide systems. This method has been approximately applied to extract the EBSD functions of Fe-pnictides and has yielded some interesting results \cite{yang:2009a,wu:2010,hwang:2015}. However, a new method has been proposed using a reverse form of the usual analytical process \cite{hwang:2015a} and has been applied to obtain the EBSD functions from the measured optical spectra of correlated multiband Fe-pnictide systems \cite{hwang:2016}. We call this new method the two-parallel-transport-channel approach.

In this study, we investigated the doping-dependent SC properties of K-doped BaFe$_2$As$_2$ (Ba-122) single crystals at four doping levels using infrared spectroscopy. Infrared/optical spectroscopy has been previously used to obtain quantitative information on the electronic band structures and charge carrier dynamics of various material systems \cite{basov:2011}. Therefore, through infrared/optical spectroscopy, the SC energy gap, spectral weight redistributions, and correlations between charge carriers (or the EBSD function) in a correlated SC material can be studied. In addition, critical physical quantities for understanding superconductivity can be extracted, including the superfluid plasma frequency (or superfluid density), London penetration depth, correlation strength, SC coherence length, SC transition temperature, and others \cite{tinkham:1975,basov:2005,carbotte:2011,hwang:2021}. We obtained the superfluid plasma frequencies and corresponding London penetration depths from the measured optical conductivity spectra of K-doped Ba-122 samples over a wide range of doping levels, from under- to overdoped. To analyze measured optical spectra of the K-doped Ba-122 samples, we employed the two-parallel-transport-channel approach \cite{hwang:2016} and obtained the EBSD functions in the SC state ($T$ = 8 K). From the extracted EBSD function, we obtained various critical physical quantities such as the coupling strength, maximum possible SC transition temperature, and SC coherence length\cite{hwang:2008c,hwang:2011,hwang:2021}. The maximum SC transition temperatures ($T_c^{\mathrm{Max}}$) estimated using the generalized McMillan formula \cite{mcmillan:1968} are greater than the actual $T_c$ measured by the DC transport technique. Therefore, the obtained EBSD functions are sufficiently large for superconductivity. We compared the doping-dependent SC properties of K-doped Ba-122 with those of hole-doped cuprates. We observed similarities and differences between the two high-temperature SC systems. These results will provide helpful information in elucidating the microscopic pairing mechanisms for both Fe-pnictide superconductors and cuprates.

\section{Results and discussion}

\subsection{Reflectance and Kramers-Kronig analysis}

\begin{figure}[!htbp]
  \vspace*{-0.3 cm}%
  \centerline{\includegraphics[width=5 in]{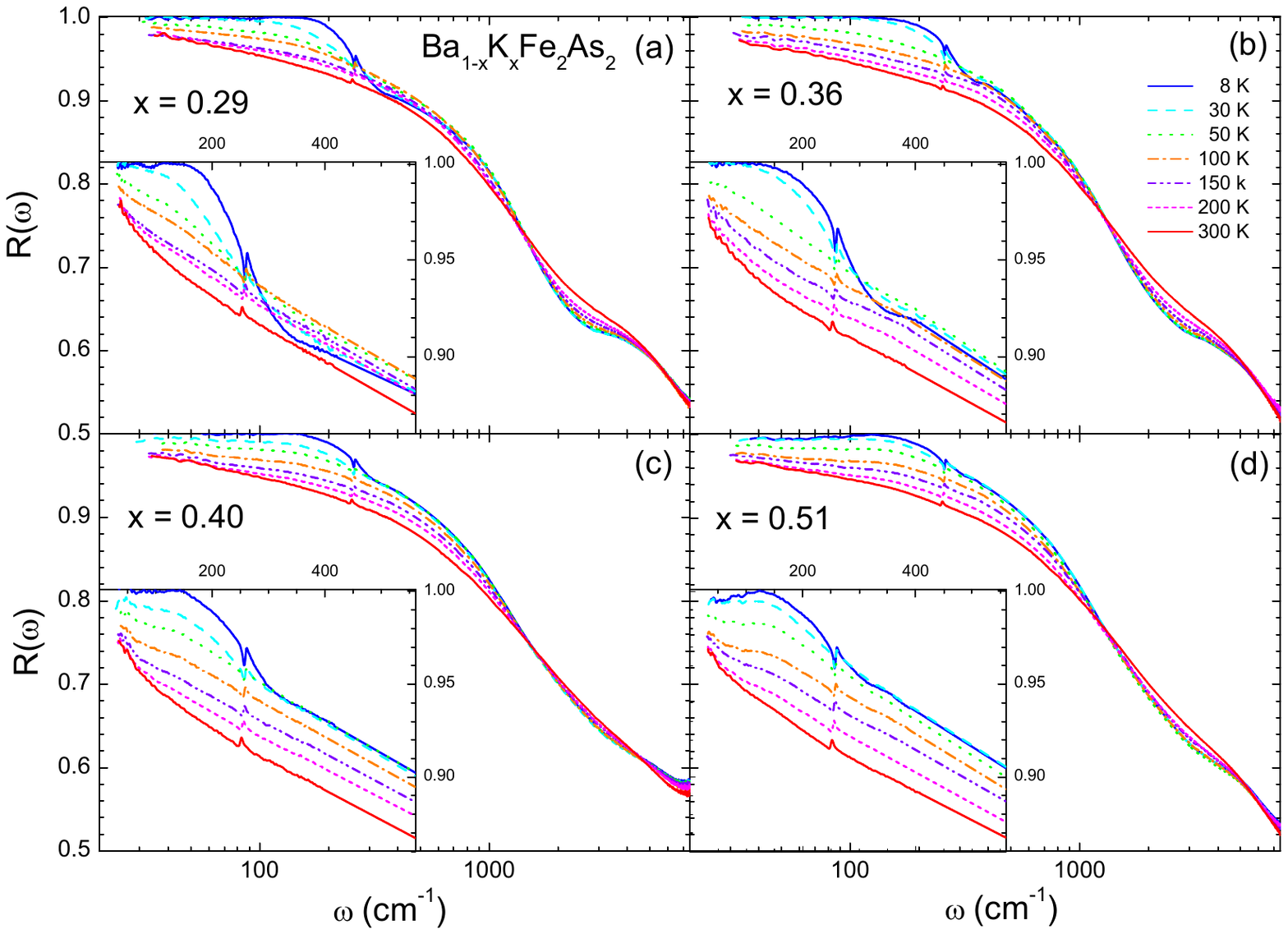}}%
  \vspace*{-0.3 cm}%
\caption{(Color online) Reflectance spectra of Ba$_{1-x}$K$_x$Fe$_2$As$_2$ single-crystal samples at four doping levels and at various selected temperatures between 8 and 300 K; (a) $x =$ 0.29, (b) $x =$ 0.36, (c) $x =$ 0.40, and (d) $x =$ 0.51.}
 \label{fig1}
\end{figure}

High-quality K-doped Ba-122 (Ba$_{1-x}$K$_x$Fe$_2$As$_2$) single crystals with four different hole-doping levels ($x =$ 0.29, 0.36, 0.40, and 0.51) were grown using a self-flux technique\cite{nakajima:2014}. The K-doped Ba-122 samples are known to have the cleanest FeAs planes among Fe-pnictide systems. The SC transition temperatures ($T_c$) of the four samples with $x =$ 0.29, 0.36, 0.40, and 0.51 are 35.9 K, 38.5 K, 38.5 K, and 34.0 K, respectively, which were determined by DC transport measurements (see Fig. S1 in the Supplementary Material). The sample with $x =$ 0.40 was optimally doped. We measured the reflectance spectra (35 - 8000 cm$^{-1}$) of the single-crystal samples at various selected temperatures below and above $T_c$ between 8 and 300 K using an {\it in-situ} gold evaporation technique \cite{homes:1993}. For infrared measurements, we used commercial Fourier-transform infrared spectrometers (IFS 113v and Vertex 80v, Bruker) and a continuous liquid helium flow cryostat. The measured reflectance spectra below 7500 cm$^{-1}$ are shown in Fig. \ref{fig1}. The reflectance spectra clearly show systematic doping- and temperature-dependencies. As we expected from the metallic ground state, the reflectance increased with decreasing the temperature in the low-frequency region. Magnified views below 550 cm$^{-1}$ are provided in the insets to show the systematic doping and temperature-dependencies more clearly. A sharp increase below $\sim$300 cm$^{-1}$, which is associated with the SC response, appears below $T_c$. A dip (or kink) near 350 cm$^{-1}$ indicates strong doping and temperature dependency, which becomes less pronounced as the doping increases at 8 K and disappears at high temperatures above 50 K. If one looks carefully at the reflectance spectra of $x$ = 0.36 and 0.51 samples near 350 cm$^{-1}$, a wiggle may be observed. This feature is not intrinsic because it comes from the merging of the far-infrared and mid-infrared spectra. There was a slight slope difference between the two measured spectral regions. It is worth noting that the $x =$ 0.51 sample shows a down-turn feature below 120 cm$^{-1}$, which is associated with disorder in the FeAs layer.

\begin{figure}[!htbp]
  \vspace*{-0.3 cm}%
  \centerline{\includegraphics[width=5 in]{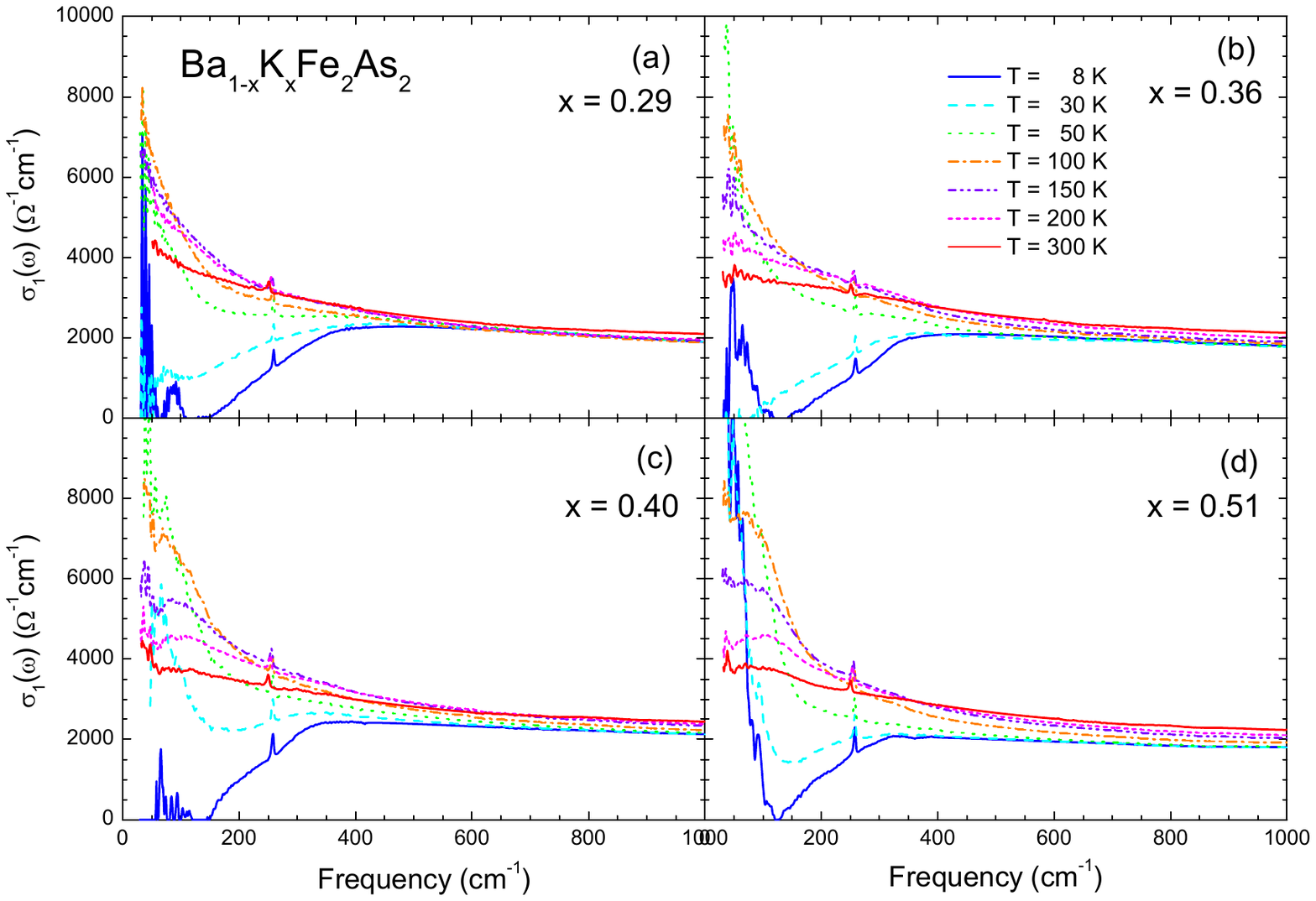}}%
  \vspace*{-0.3 cm}%
\caption{(Color online) Optical conductivity spectra of Ba$_{1-x}$K$_x$Fe$_2$As$_2$ single crystal samples at four doping levels and at various selected temperatures between 8 and 300 K; (a) $x =$ 0.29, (b) $x =$ 0.36, (c) $x =$ 0.40, and (d) $x =$ 0.51.}
 \label{fig2}
\end{figure}

The optical conductivity spectra were obtained from the measured reflectance spectra using Kramers-Kronig analysis \cite{wooten,tanner:2019}. To perform this analysis, the measured reflectance in a finite spectral range must be extrapolated to zero and infinity. For extrapolation to zero frequency, we used the Hagen-Rubens relation ($1-R(\omega) \propto \sqrt{\omega}$) for the normal state and $1-R(\omega) \propto \omega^4$ for the superconducting state. For extrapolation to infinity, we used available published data \cite{dai:2013a} up to 40000 cm$^{-1}$ and assumed that $R(\omega) \propto \omega^{-1}$ from 40000 to 10$^6$ cm$^{-1}$ and $R(\omega) \propto \omega^{-4}$ (i.e., the free-electron response) above 10$^6$ cm$^{-1}$. Fig. \ref{fig2} shows the real parts of the optical conductivities of the four K-doped Ba-122 samples at various selected temperatures. We observed clear doping- and temperature-dependent trends. As the temperature decreased to $T_c$, the conductivity at near zero frequency monotonically increased due to the reduction of the scattering rate. The temperature-dependent trends of typical and correlated metals in the low frequency region are not monotonic due to the thermally excited phonon contribution. As an example, for the Drude metal, as the temperature decreases, whereas the spectral weight of the Drude mode (or charge carrier density) is conserved, the width of the Drude mode decreases, resulting in a higher DC conductivity. As the doping increased, the temperature-dependent change was enhanced. The conductivity was significantly suppressed below $\sim$ 350 cm$^{-1}$ below $T_c$, because of the SC gap formation. The strong absorption of the sample with $x =$ 0.51 at 8 K below $\sim$ 200 cm$^{-1}$ (or inside the SC gap) may have been associated with localization effects caused by a disorder in the two-dimensional FeAs layer \cite{singley:2001}. The absorption features of the other three samples at 8 K below 200 cm$^{-1}$ might be regarded artifacts due to unavoidable experimental uncertainties in the reflectance near the perfect reflectance of 1.0.

\subsection{Superconducting properties}

\begin{figure}[!htbp]
  \vspace*{-0.3 cm}%
  \centerline{\includegraphics[width=5 in]{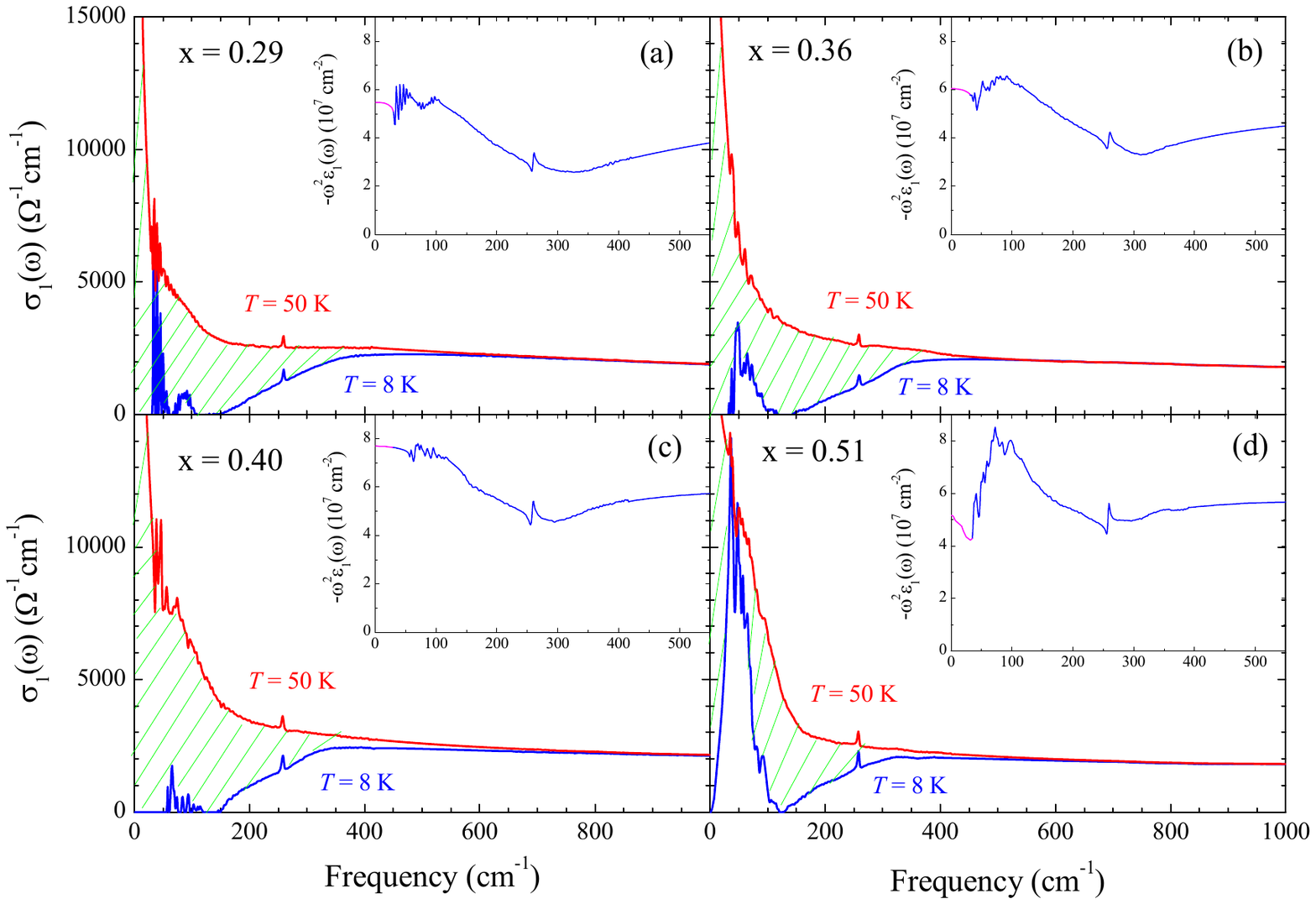}}%
  \vspace*{-0.3 cm}%
\caption{(Color online) Optical conductivity spectra of K-doped Ba-122 at two temperatures: 8 K (SC state) and 50 K (normal state); (a) $x =$ 0.29, (b) $x =$ 0.36, (c) $x =$ 0.40, and (d) $x =$ 0.51. In the insets, $-\omega^2\epsilon_1(\omega)$ as functions of $\omega$ are shown.}
 \label{fig3}
\end{figure}

We focused only on the doping-dependent SC properties of the K-doped Ba-122 samples. We estimated the superfluid plasma frequencies of all four samples using two independent methods: one from the real part of the optical conductivity ($\sigma_1(\omega)$) and the other from the imaginary part ($\sigma_2(\omega)$). In the first method, when the material system transitions from the normal to the SC state, a fraction of the normal charge carriers become superfluid carriers. Because the scattering rate of the superfluid carriers is essentially zero, their spectral weight is piled up (i.e., appears as a delta function) at zero frequency. Therefore, the superfluid spectral weight seems to disappear from the finite-frequency region. The apparently absent spectral weight is called the missing spectral weight. The superfluid plasma frequency can be estimated from this missing spectral weight. Fig. \ref{fig3} shows the optical conductivity spectra at 8 K ($< T_c$) and 50 K ($> T_c$). The missing spectral weights for all four samples are marked with the hatched areas. The apparent missing spectral weight can be used to calculate the superfluid plasma frequency ($\Omega_{sp}$), as $\Omega_{\mathrm{sp}}^2 = (120/\pi)\int_{0^+}^{\infty}[\sigma_{1,N}(\omega, 50 K)-\sigma_{1,SC}(\omega, 8 K)] d\omega$, where $\sigma_{1,N}(\omega, 50 K)$ and $\sigma_{1,SC}(\omega, 8 K)$ are the real parts of the optical conductivity in the normal state (50 K) and SC state (8 K), respectively. Here, the units of all frequencies are cm$^{-1}$. This method is also called the Ferrell–Glover–Tinkham (FGT) sum rule \cite{glover:1956,ferrell:1958}. In our study, the estimated superfluid plasma frequencies ($\Omega_{\mathrm{sp}}$) were 7422$\pm$400, 7518$\pm$300, 8418$\pm$500, and 6584$\pm$400 cm$^{-1}$ for $x =$ 0.29, 0.36, 0.40, and 0.51, respectively. It is worth noting that, in principle, if the optical conductivity data at the same temperature for the superconducting and normal states were used for the FGT sum rule, the best result is expected. However, because, practically, we cannot obtain the optical conductivity at 8 K for the normal state, we used the optical conductivity at our lowest temperature for the normal state. This approximation is usually used for estimating the superfluid plasma frequency using the FGT sum rule. In the second method, the imaginary part of the optical conductivity (or the real part of the dielectric function $\epsilon_1(\omega)$) can be used to estimate the superfluid plasma frequency. The imaginary part of the optical conductivity ($\sigma_2(\omega)$) of the superfluid carriers rapidly increases in the low-frequency region and diverges at zero frequency, that is, $\sigma_2(\omega)=\Omega_{\mathrm{sp}}^2/4\pi \omega=(\omega/4\pi)[1-\epsilon_1(\omega)]$. As a result, $\Omega_{\mathrm{sp}}^2 = \lim_{\omega\rightarrow 0}[-\omega^2\epsilon_1(\omega)]$. $-\omega^2\epsilon_1(\omega)$ is shown as a function of $\omega$ in the insets of Fig. \ref{fig3}. In our study, the estimated superfluid plasma frequencies ($\Omega_{\mathrm{sp}}$) obtained from $-\omega^2\epsilon_1(\omega)$ were 7396, 7776, 8766, and 7178 cm$^{-1}$ for $x =$ 0.29, 0.36, 0.40 and 0.51, respectively. The corresponding superfluid plasma frequencies obtained using the two methods were in good agreement. The uncertainties of the superfluid plasma frequencies were obtained by considering experimental uncertainty ($\pm$0.5\%) in the measured reflectance. It is worth noting that the method using the imaginary part of the optical conductivity experiences small uncertainties because the reflectance becomes 1.0 below the SC gap for $s$-wave superconductors. As mentioned previously, the $x =$ 0.51 sample shows a strong absorption inside the SC gap. This absorption may suppress the extracted superfluid density (see Fig. S2 in the Supplementary Material). The London (or magnetic field) penetration depth ($\lambda_{\mathrm{L}}$) is one of the two characteristic length scales of superconductivity. The other length scale is the SC coherence length, which is directly associated with the size of the Cooper pairs. Later in this paper, we will discuss how we obtained the coherence length from the measured infrared spectrum. The London penetration depth is directly related to the superfluid plasma frequency as $\lambda_{\mathrm{L}} = 1/(2 \pi \Omega_{\mathrm{sp}})$. In this study, the estimated $\lambda_{\mathrm{L}}$ were 2144 (2152), 2117 (2047), 1891 (1816), and 2417 (2217) \AA $\:$ for $x =$ 0.29, 0.36, 0.40 and 0.51, respectively. The values in parentheses are the corresponding $\lambda_{\mathrm{L}}$ values for the superfluid plasma frequencies obtained from $-\omega^2\epsilon_1(\omega)$. The optimally doped sample exhibited the shortest London penetration depth, indicating that its superconductivity was the strongest. These estimated London penetration depths are comparable to those of other doped Ba-122 Fe-pnictides \cite{li:2008,heumen:2010,nakajima:2010,kim:2010,yoon:2017}. The London penetration depth of Bi$_2$Sr$_2$CaCu$_2$O$_{8+\delta}$ (Bi-2212) exhibited a different doping-dependent trend; it decreased monotonically with increasing the hole doping level \cite{hwang:2007a}.

\begin{figure}[!htbp]
  \vspace*{-0.3 cm}%
  \centerline{\includegraphics[width= 5 in]{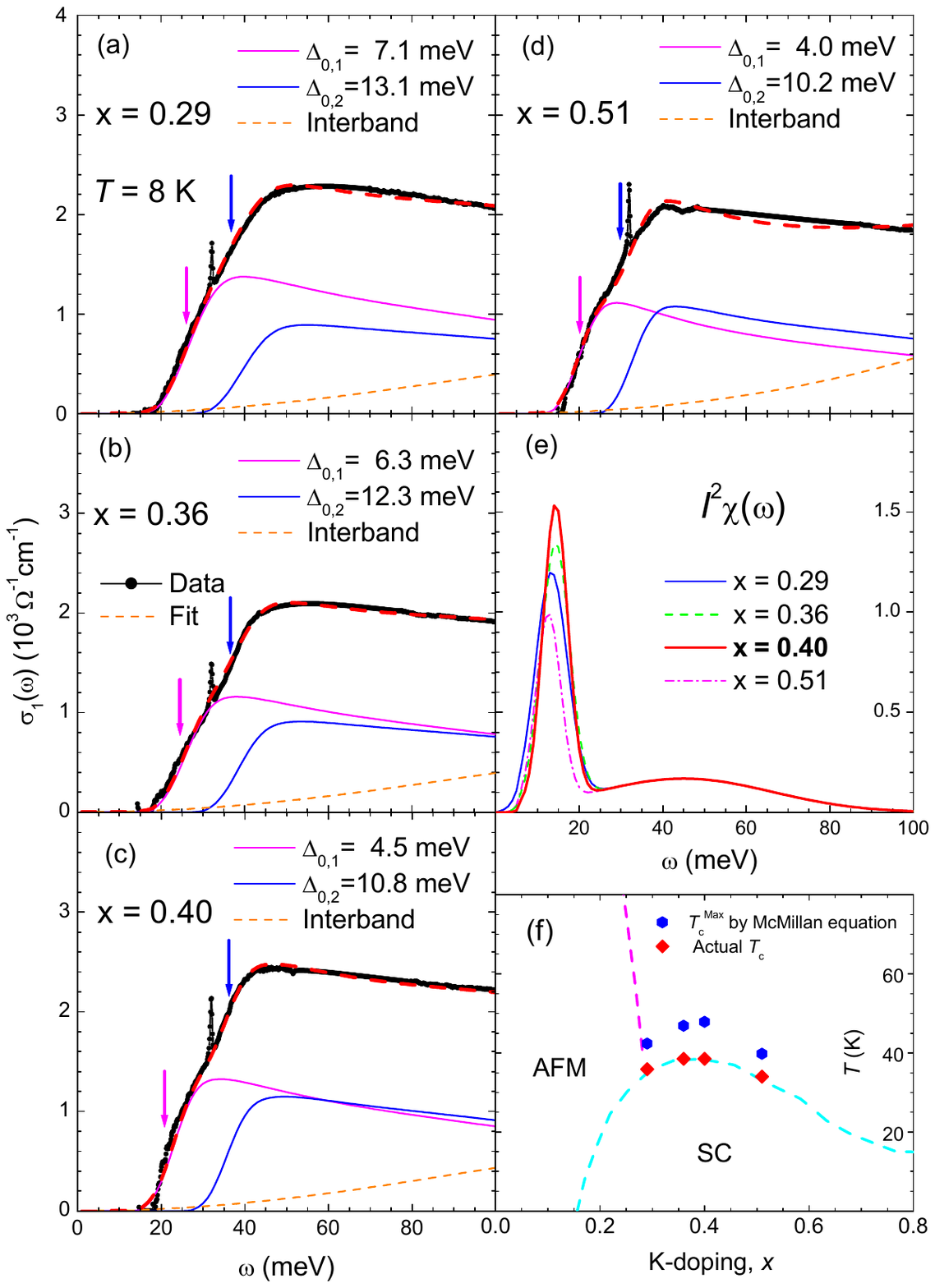}}%
  \vspace*{-0.3 cm}%
\caption{(Color online) (a-d) Optical conductivity spectra of K-doped Ba-122 at 8 K (SC state) and their fits; (a) $x =$ 0.29, (b) $x =$ 0.36, (c) $x =$ 0.40, and (d) $x =$ 0.51. (e) Resulting electron-boson spectral density functions ($I^2\chi(\omega)$). (f) Maximum SC transition temperature ($T_c^{\mathrm{Max}}$) and actual measured $T_c$. The phase diagram of K-doped Ba-122 was adopted from published literature \cite{nakajima:2014}.}
 \label{fig4}
\end{figure}

The two-parallel-transport-channel approach was previously proposed to extract the EBSD function from the measured infrared conductivity of multiband correlated electron systems, Fe-pnictides \cite{hwang:2016}. We used the proposed approach to obtain the EBSD functions of our four K-doped Ba-122 samples in the SC state (8 K). Figs. \ref{fig4} (a)-(d) show the optical conductivity spectra (black symbols) and fits (thick red dashed lines) of all four K-doped samples ($x =$ 0.29, 0.36, 0.40, and 0.51). Two SC gaps ($\Delta_{0,1}$ and $\Delta_{0,2}$), and two corresponding plasma frequencies ($\Omega_{P,1}$ and $\Omega_{P,2}$) for the two transport channels, and two (sharp and broad) Gaussian peaks for the input EBSD function ($I^2\chi(\omega)$) were required to obtain these fits. Here, $I$ is the coupling constant between an electron and the mediated bosons, and $\chi(\omega)$ is the spectrum of the mediated boson. We note that the broad peak shares the same origin as the sharp one and plays a role in superconductivity. The broad peak may also depend on doping, even though its doping dependence will be negligibly small compared with the sharp peak. Therefore, we approximately assumed that the broad peak was doping-independent for the analysis. As in the previously published paper\cite{hwang:2016}, the EBSD functions of two channels are also assumed to be the same. In fact, considering that an optically measured spectrum is $k$-space averaged, the assumption is reasonable because the EBSD function of each channel cannot be separated from the measured optical spectrum and can be a $k$-space averaged one. For the two-parallel-transport-channel approach, the assumed EBSD function can be the averaged EBSD function of the two channels because each channel is originated from a band in a different k-space at the Fermi level. We also assumed that the impurity scattering rates for the two channels were zero because this system is known to be a clean system \cite{hwang:2016}. In the optimally K-doped Fe-pnictides \cite{christianson:2008,hwang:2016}, we fixed the position ($\Omega_R$) of the sharp Gaussian peak at $\sim4.3 k_B T_c$, as in the previously reported magnetic resonance mode. Because it is nearly independent of doping, the same broad Gaussian peak was used for all four samples, as shown in Fig. \ref{fig4}(e). We required an interband transition (orange dashed line) located in the low-frequency region \cite{benfatto:2011} for each doping. As the doping increased, the sizes of both SC gaps ($\Delta_{0,1}$ and $\Delta_{0,2}$) monotonically decreased, as shown in the figure. This was the same trend observed on the overdoped side of the hole-doped Bi-2212\cite{hwang:2007}. The SC gaps of the optimally doped sample were consistent with the SC gaps observed in an angle-resolved photoemission (ARPES) study \cite{ding:2008}. We clearly observed two sharp increases (marked by arrows) in the optical conductivity, the energy scales of which were $\Omega_R + 2\Delta_{0,i}$, where $i =$ 1 or 2. The two sharp increases can be seen from the slope change in the conductivity (see Fig. S3(Left) in the Supplementary Material). Note that, with one band alone, one cannot properly fit the measured spectrum (see Fig. S3(Right) in the Supplementary Material). It should be noted that a previous optical study was conducted to obtain the EBSD function of a Co-doped Ba-122 for a normal state using a single band alone \cite{wu:2010}. The resulting EBSD function consists of two components. The plasma frequencies ($\Omega_{P,1}$ and $\Omega_{P,2}$) for the two channels were 11900 and 10750 cm$^{-1}$, 10800 and 10900 cm$^{-1}$, 11400 and 11800 cm$^{-1}$, and 9750 and 11000 cm$^{-1}$ for $x =$ 0.29, 0.36, 0.40, and 0.51, respectively. As the doping increased, the ratio of the spectral weight ($\equiv \Omega_{P,1}^2 \pi/120$) of the small SC gap ($\Delta_{0,1}$) to that ($\equiv \Omega_{P,2}^2\pi/120$) of the large SC gap ($\Delta_{0,2}$) monotonically decreased. We also estimated the fraction of charge carriers condensed into the superfluid from the fitting parameters and superfluid plasma frequencies, that is, $\Omega_{sp}^2/(\Omega_{P,1}^2+\Omega_{P,2}^2)$. The estimated fractions were 0.21, 0.26, 0.29, and 0.24 for $x =$ 0.29, 0.36, 0.40, and 0.51, respectively, indicating that the condensation rate per charge carrier was maximized in the optimally doped sample.

The resulting extracted $I^2\chi(\omega)$ of all four samples are shown in Fig. \ref{fig4}(e). The amplitude of the sharp Gaussian peak for the optimally doped sample was the largest. From the resulting $I^2\chi(\omega)$, we estimated the coupling constant ($\lambda$), which is defined by $\lambda \equiv 2 \int_0^{\omega_c} I^2\chi(\omega')/\omega' \: d \omega'$, where $\omega_c$ is the cutoff frequency (for our case, 100 meV). The estimated coupling constants were 2.34, 2.24, 2.11, and 1.68 for $x =$ 0.29, 0.36, 0.40, and 0.51, respectively. The coupling constant monotonically decreased as the doping increased, which was similar to the doping-dependent trend of the coupling constant of hole-doped Bi-2212 \cite{hwang:2007,hwang:2007a}, supporting the microscopic SC mechanism mediated by antiferromagnetic spin fluctuations. It worth noting that the doping-dependent trend of the coupling constant was consistent with that of K-doped Ba-122 at 50 K obtained including the pseudogap in the analysis \cite{lee:2022}. The SC transition temperature ($T_c$) can be estimated from $I^2\chi(\omega)$ using the generalized McMillan's equation $k_B T_c \cong 1.13 \: \hbar\omega_{\mathrm{ln}}\exp{[-(1+\lambda)/(g \lambda)]}$, where $k_B$ is the Boltzmann constant, $\hbar$ is the reduced Planck constant, $g$ is an adjustable parameter between 0.0 and 1.0 that opposes superconductivity, and $\omega_{\mathrm{ln}}$ is the logarithmically averaged frequency of $I^2\chi(\omega)$, that is, $\omega_{\mathrm{ln}} \equiv \exp{[(2/\lambda) \int_0^{\omega_c}\ln{\omega'} \: I^2\chi(\omega')/\omega' \: d\omega']}$. When $g$ = 1.0, $T_c$ reaches its maximum value, that is, $T_c^{\mathrm{Max}}$. The estimated $T_c^{\mathrm{Max}}$ values were 42.4, 46.9, 47.9, and 39.8 K for $x =$ 0.29, 0.36, 0.40, and 0.51, respectively. Each $T_c^{\mathrm{Max}}$ (blue hexagon) was greater than the actual $T_c$ (red diamond) measured by the DC transport technique, indicating that the extracted $I^2\chi(\omega)$ was sufficiently strong for the superconductivity, as shown in Fig. \ref{fig4}(f). The $T_c^{\mathrm{Max}}$ peaked at optimal doping, which was the same doping-dependent trend as the $T_c^{\mathrm{Max}}$ of hole-doped cuprates\cite{hwang:2021}.

The extracted $I^2\chi(\omega)$ also contains information on the characteristic timescale of the retarded interaction between charge carriers \cite{hwang:2021}. The SC coherence length ($\xi_{\mathrm{SC}}$) can be estimated from the timescale and Fermi velocity ($v_{\mathrm{F}}$). The SC coherence length is the other length scale of the two characteristic length scales for superconductivity, which is intimately associated with the size of the Cooper pair. The average frequency of $I^2\chi(\omega)$ is defined as $\langle \Omega \rangle \equiv (2/\lambda) \int_0^{\omega_c} \omega' \: [I^2\chi(\omega')/\omega']\: d\omega'$ and is directly related to the characteristic timescale of the retarded interaction. The estimated average frequencies ($\langle \Omega \rangle$) were 16.80, 18.44, 18.73, and 18.72 meV for $x =$ 0.29, 0.36, 0.40, and 0.51, respectively. The SC correlation length ($\xi_{\mathrm{SC}}$) can be expressed in terms of $\langle \Omega \rangle$ and $v_{\mathrm{F}}$ as $\xi_{\mathrm{SC}} = (1/2) v_{\mathrm{F}} [2 \pi/\langle \Omega \rangle] (1/2\pi) = (1/2) v_{\mathrm{F}}/\langle \Omega \rangle$ \cite{hwang:2021}. Here, we assumed isotropic SC gaps, spin-singlet pairings, and force-mediated bosons resulting from the antiferromagnetic spin fluctuations. Note that the factor of $1/2\pi$ originates from the two-dimensional character of the charge transport and the prefactor of 1/2 results from our assumptions of the spin-singlet pairing and antiferromagnetic spin fluctuations \cite{hwang:2021}. The average Fermi velocity of the optimally doped sample ($x =$ 0.40) was approximately 0.38 eV\AA $\:$ as derived from ARPES \cite{ding:2011}. The estimated SC coherence lengths ($\xi_{\mathrm{SC}}$) were 11.31, 10.30, 10.14, and 10.15 \AA $\:$ for $x =$ 0.29, 0.36, 0.40, and 0.51, respectively. We assumed that the Fermi velocity was the same for all four doping levels. The optimally doped sample exhibited the shortest coherence length, which showed the same doping-dependent trend as that of hole-doped cuprates \cite{hwang:2021}. The estimated SC coherence length was consistent with the reported values of the optimally K-doped sample \cite{welp:2009,ding:2011}. The doping-dependent SC physical quantities are summarized in Table 1. The table shows that the ratios $\lambda_{\mathrm{L}}$ to $\xi_{\mathrm{SC}}$ are very large ($\sim$ 200), indicating that the SC material (K-doped Ba-122) systems are type-II superconductors.

\begin{table}[h!]
\begin{tabular}{|c||c|c|c|c|c|}
  \hline
  \mbox{Doping} ($x$) & $T_c$ (K) & $\lambda$ & $T_c^{\mathrm{Max}}$ (K) & $\lambda_{\mathrm{L}}$ (nm)& $\xi_{\mathrm{SC}}$ (\AA) \\
  \hline\hline
  0.29 & 35.9 & 2.34 & 42.4 & 215.2 & 11.31 \\  \hline
  0.36 & 38.5 & 2.24 & 46.9 & 204.7 & 10.30 \\  \hline
  0.40 & 38.5 & 2.11 & 47.9 & 181.6 & 10.14 \\  \hline
  0.51 & 34.0 & 1.68 & 39.8 & 221.7 & 10.15 \\
  \hline
\end{tabular}
\caption{Doping-dependent SC physical quantities of K-doped Ba-122 (Ba$_{1-x}$K$_x$Fe$_2$As$_2$).}
\label{table:1}
\end{table}

\section{Conclusions}

We obtained various SC physical quantities from measured infrared spectra of K-doped Ba-122 single crystals in a wide doping range from under- to overdoped levels. The superfluid plasma frequencies ($\Omega_{\mathrm{sp}}$) were determined using two independent methods. The doping-dependent $\Omega_{\mathrm{sp}}$ peaked at the optimal doping level, which was different from the doping-dependent trend of hole-doped Bi-2212 cuprates \cite{hwang:2007a}, where the $\Omega_{\mathrm{sp}}$ monotonically decreased as the doping increased. The EBSD functions were obtained using a two-parallel-transport-channel approach \cite{hwang:2016}. We determined the doping-dependent trends of the two crucial characteristic lengths of the superconductivity ($\lambda_{\mathrm{L}}$ and $\xi_{\mathrm{SC}}$). $\lambda_{\mathrm{L}}$ was derived from the superfluid plasma frequency, whereas $\xi_{\mathrm{SC}}$) was derived from the characteristic timescale of the SBSD function and reported Fermi velocity. The estimated characteristic SC lengths ($\lambda_{\mathrm{L}}$ and $\xi_{\mathrm{SC}}$) were consistent with the values reported by other experimental techniques. The ratio $\lambda_{\mathrm{L}}$ to $\xi_{\mathrm{SC}}$ was quite large ($\sim$ 200), indicating that the materials are type-II superconductors. The coupling constants ($\lambda$) were derived from the extracted EBSD functions. The coupling constant exhibited a monotonic decrease as the doping increased, similar to the doping-dependent trend of $\lambda$ for hole-doped cuprates \cite{hwang:2007,hwang:2007a}. The maximum SC transition temperatures ($T_c^{\mathrm{Max}}$) estimated from the EBSD functions were greater than those measured by the DC transport technique, suggesting that the coupling constant was sufficiently strong to explain the superconductivity. The doping-dependent $T_c^{\mathrm{Max}}$ showed the largest value at the optimal doping level, exhibiting the same doping-dependent trend as the $T_c^{\mathrm{Max}}$ of hole-doped cuprates\cite{hwang:2021}. The obtained doping-dependent SC quantities of K-doped Ba-122 exhibited similarities and differences from those of hole-doped cuprates \cite{hwang:2007a,hwang:2021}. Our results may be beneficial in establishing the microscopic pairing mechanism of high-temperature superconductors (Fe-pnictides and cuprates).
\\ \\
\noindent {\bf Acknowledgements} This study was supported by the National Research Foundation of Korea (NRFK Grant Nos. 2020R1A4A4078780, and 2021R1A2C101109811).
\\ \\

%
%
\bibliographystyle{naturemag}
\bibliography{bib}

\end{document}